# Searching for topological Fermi arcs via quasiparticle interference on a type-II Weyl semimetal MoTe$_2$

Davide Iaia[1], Guoqing Chang[2], Tay-Rong Chang[3], Jin Hu[4], Zhiqiang Mao[4], Hsin Lin[2], Shichao Yan[5*] & Vidya Madhavan[1*]

[1] *Department of Physics and Frederick Seitz Materials Research Laboratory, University of Illinois Urbana-Champaign, Urbana, Illinois 61801, USA*
[2] *Institute of Physics, Academia Sinica, Taipei 11529, Taiwan*
[3] *Department of Physics, National Cheng Kung University, Tainan 701, Taiwan*
[4] *Department of Physics and Engineering Physics, Tulane University, New Orleans, Louisiana 70018, USA*
[5] *School of Physical Science and Technology, ShanghaiTech University, Shanghai, 201210, China*
[*]*Emails*: yanshch@shanghaitech.edu.cn; vm1@illinois.edu

**Weyl semimetals display a novel topological phase of matter where the Weyl nodes emerge in pairs of opposite chirality and can be seen as either a source or a sink of Berry curvature. The exotic effects in Weyl semimetals, such as surface Fermi arcs and the chiral anomaly, make them a new playground for exploring novel functionalities. Further exploiting their potential applications requires clear understanding of their topological electronic properties, such as Weyl points and Fermi arcs. Here we report a Fourier transform scanning tunneling spectroscopy (FT-STS) study on a type-II Weyl semimetal candidate MoTe$_2$ whose Weyl points are predicated to be located above Fermi level. Although its electronic structure below the Fermi level have been identified by angle resolved photo emission spectroscopy (ARPES), by comparing our experimental data with first-principles calculations, we are able to identify the origins of the multiple scattering channels at energies both below and above Fermi level. Our calculations also show the existence of both trivial and topological arc like states above the Fermi energy. In the FT-STS experiments, we have observed strong signals from intra-arc scatterings as well as from the scattering between the arc-like surface states and the projected bulk states. A detailed comparison between our experimental observations and calculated results reveals the trivial and non-trivial scattering channels are difficult to distinguish in this compound. Interestingly, we find that the broken inversion symmetry changes the terminating states on the two inequivalent surfaces, which in turn changes the relative strength of the scattering channels observed in the FT-STS images on the two surfaces.**

Transition metal dichalcogenides (TMDs) provide a fertile ground for exploring exotic collective quantum phenomena, such as superconductivity and charge density waves[1-3]. Recently, the $Mo_xW_{1-x}Te_2$ class of TMDs has attracted great interest due to the topological nature of the electronic states and as a topological Weyl semimetal (TWS)[4-7]. Their low-energy excitations behave as Weyl fermions that always appear in pairs with opposite chirality. TWSs are classified into type-I and type-II: type-I TWSs have point like Fermi surfaces and respect Lorentz invariance, while type-II with tilted Weyl cones break Lorentz invariance[5,8,9]. Type-I TWSs were first predicted and experimentally discovered in the (Ta, Nb) (As, P) family compounds[8-11]. Type-II TWSs have been proposed to exist in the $Mo_xW_{1-x}Te_2$ family TMDs[4-7] and LaAlGe[12]. In this system, the tilted Weyl cones that arise from topologically protected crossings of valence and conduction bands cause touching points between electron and hole pockets near the Fermi level. Exotic Fermi arcs states that join these touching points, are expected on the surfaces of these materials. In the last few years, there have been a series of experimental studies aimed at verifying the existence of the Fermi arcs[13-21].

For the $Mo_xW_{1-x}Te_2$ family, the Weyl nodes are usually located at energies above the Fermi level which makes it difficult to be accessed by conventional angle resolved photo emission spectroscopy (ARPES)[13-15,17,20]. Fourier transform scanning tunneling spectroscopy (FT-STS) contains information on the scattering vectors between electronic states in momentum space and can provide the information on both the occupied and unoccupied states[22] which makes it a particularly powerful probe for the topological states of the $Mo_xW_{1-x}Te_2$. There have been several FT-STS studies on the $Mo_xW_{1-x}Te_2$[15,23-28]. However, due to the complex band structure and the low signal strength in the FT-STS images, the existence of Fermi arcs in $Mo_xW_{1-x}Te_2$ system still remains to be clarified.

Here we use low-temperature scanning tunneling microscopy (STM) and first-principles calculations to directly visualize and identify the electronic states in $MoTe_2$, an end member of the $Mo_xW_{1-x}Te_2$ system. The other end member, $WTe_2$ was the first predicated type-II Weyl semimetal, but the small momentum separation between the opposite Weyl points that are located about 50 mV above the Fermi level, made experimental confirmation by conventional ARPES difficult [5,13]. $MoTe_2$ was later predicated as another candidate of type-II Weyl semimetal where the Weyl points have six times larger momentum separation which makes experimental probe of Weyl nodes much easier[6,7]. For $MoTe_2$, previous ARPES works have found evidence for the trivial and topologically protected Fermi arcs at energies below Fermi level[18-20]. However, for various reasons such as the size of the STS maps or the properties of the scattering impurities, previous FT-STSs on $MoTe_2$ did not show sufficient signal strength to perform a clear analysis of the FT-STS data, and have not arrived at an unequivocal conclusion about the nature of the Weyl states. In this work, we present FT-STS data with much higher signal strength that allows us to compare the various scattering channels observed in the experimental data with theory, and reach a robust conclusion on the origins of the scattering channels. Interestingly, based on this higher resolution FT-STS data, we come to different conclusions compared to the previous work[15,23,24,26]: the trivial and non-trivial scattering channels are indistinguishable in the FT-STS images. Moreover, the broken inversion symmetry in this compound is reflected in the FT-STS images taken on two inequivalent surfaces.

**Results**

The $T_d$ phase crystal has an orthorhombic structure with van de Waals stacking of Te-Mo-Te sandwich layers along the *c*-axis direction as shown in Fig. 1a. Due to the weak van de Waals interaction between the Te layers, the MoTe$_2$ sample cleaves at the Te layers and the cleaved surface is Te-terminated. The bulk, and surface Brillouin zones for this termination, are shown in Fig. 1b. In high-resolution STM topographies, the two inequivalent Te atomic rows in the top Te layer can be clearly resolved (Fig. 1d). There are mainly two kinds of atomic defects and they look different in the STM topographies taken with positive bias voltages (as shown in the red and black circles in Fig. 1e). The defects lie on the Te sites and are likely to be Te vacancies/impurities in the two Te atomic rows at the top Te layer. In addition, in the negative bias topography, there are other faint defects that may be attributed to vacancies/impurities in the layers underneath (more details can be found in Supplementary Fig. S1 and Fig. S2).

Fig. 1f shows the position dependent d*I*/d*V* spectra taken on a 50 nm by 50 nm area on MoTe$_2$ in the energy range of ±500 mV. In general, the d*I*/d*V* spectra are parabola-shaped with minimum around 10 mV above the Fermi level. Despite the presence of impurities, the low energy density of states as well as the overall parabola-shape of the spectra remains homogeneous and there are no clear impurity-induced resonance states in this energy range[26]. However, the scattering pattern of quasiparticles by the atomic impurities can be clearly seen in the d*I*/d*V* maps, Fig. 1c. The next step is to study the Fourier transforms of the d*I*/d*V* maps to look for signatures of the electronic states in momentum space.

Figs. 2 a-i show FT-STS images at a few energies below and above the Femi level (also see Figs. S6 and S7). There is a distinct change in the FT-STS as we go from negative to positive energies. With increasing the bias above the Fermi level, several strong scattering vectors gradually emerge. To identify the origin of the scattering vectors we carried out first-principles density functional theory (DFT) calculations of band structures of MoTe$_2$ which were then used to calculate FT-STS images. Using the iterative Green's function method, we obtained the surface states of MoTe$_2$ under different terminations (details in Supplementary Section 8 which were then used to simulate the FT-STS patterns using spin-dependent scattering probability method).

As can be seen in Fig. 3, the simulated FT-STS matches well with the measured FT-STS over a wide range of energies. In the measured FT-STS at energies below −100 mV (right panel in Fig. 3a), there is a lip-like feature with a bright center. By comparing with the simulated FT-STS at −140 mV, we conclude that this lip-like feature is due to the scattering of the bulk states of MoTe$_2$ (Supplementary Fig. S5). At −40 mV, the measured FT-STS shows two short arc-like features at the left or right side of the bright center spot (right panel of Fig. 3b). Due to the multiple scattering vectors seen in the simulated FT-STS at negative bias voltages, it is difficult to unambiguously identify the origin of these two short arcs. However, by comparing the momentum-space position of these short arcs, we conclude that they most likely arise from the scattering processes from the trivial surface states to the hole-pockets. At positive bias voltages, the scattering patterns become much clearer (Fig. 2i). By comparing the calculated FT-STS with the measured FT-STS at +40 mV and +100 mV (Figs. 3c and 3d and supplemental Fig. S9), we can for the first time, identify all the scattering patterns as: $Q_1$ is induced by the intra-scattering of the topologically trivial surface states; $Q_2$ is caused by the scattering between the surface states and the electron pockets. $Q_3$ and $Q_4$ are due to the scattering between the surface states and the projected bulk states at the $\bar{Y}$ point.

Although the high-resolution FT-STS data clearly show multiple scattering vectors from the surface states and our experimental measurements match the numerical calculations, these data alone do not provide unambiguous evidence for the existence of non-trivial Fermi arcs. According to our calculations, the Weyl points for MoTe$_2$ are located at 4 mV and 57 mV. As the energy increased from the Fermi level to +100 mV, the scattering vectors ($Q_1$ to $Q_4$) gradually emerge in the measured FT-STSs (Figs. 2f-i), and there is no sudden change in the FT-STSs near the energy positions of the Weyl nodes. This suggests that there are no clear features induced by the nontrivial Fermi arcs in the FT-STSs on MoTe$_2$. We note that our conclusions are in contrast with previous studies[15,23,24,26,27]. Most previous STM studies have claimed that Mo$_x$W$_{1-x}$Te$_2$ family material is a Weyl semimetal based on the existence of the long Fermi arcs in the FT-STS data. However, based on our theoretical and experimental studies we find that the long arcs arise primarily from topologically trivial scattering channels. In addition, for Mo$_{0.66}$W$_{0.34}$Te$_2$ samples, the absence of the scattering vector ($Q_3$) between the arc-like feature and projected bulk states was taken as poof of the existence of the topological Fermi arc[27]. In contrast, in our high-resolution data on MoTe$_2$, we clearly detect this scattering vector ($Q_3$) thereby determining that the scattering process from trivial surface states forms an important component of the scattering seen in our FT-STS data.

**Discussion**

This leads to the question: why is it difficult to detect the nontrivial Fermi arcs in MoTe$_2$ even with high-resolution FT-STS measurements? Weyl fermions are local singularities in momentum space (Fig. S10). Therefore, at energies or momentum space away from the Weyl nodes, the non-trivial properties induced by Weyl fermions are negligible. In figure 4 we show the calculated surface states of MoTe$_2$ at three different energies, below, at, and above the Fermi level. Figs. 4a-c show constant energy contours at −40 mV, 0 mV and +40 mV respectively. Figs. 4d-f are the zoom-in of Figs. 4a-c in the momentum region around the positions of the Weyl nodes (the orange rectangle area shown in Figs. 4a-c). Fig. 4d clearly shows that the long topologically trivial arc in Fig. 4a lies far from the Weyl nodes in momentum space, and it is therefore not related to the Weyl nodes. With increasing energy, the topologically trivial arc moves towards the Weyl nodes and starts to touch the Weyl nodes at the Fermi level (Fig. 4e). In the constant energy contour of $E$ = +40meV, the long trivial arc is disconnected with the Weyl nodes and a small topological Fermi arc (arc 2) can be seen in between the two Weyl nodes, as shown in Fig. 4f (also see Fig. S10). However, the separation in momentum space between the nontrivial arc (arc 2) and the trivial arcs (arc 1 and 3) is only about 0.008 Å$^{-1}$, which makes it extremely difficult to distinguish the scattering features in FT-STS induced by the nontrivial arc and those induced by the long trivial arc. Thus, by comparing the STM data with calculations, we conclude that the trivial and non-trivial scattering channels are indistinguishable in this compound.

Finally, our STM data show signatures of the broken inversion symmetry that's critical to the physics of MoTe$_2$. Due to the bulk broken inversion symmetry, the two surfaces obtained after cleaving are inequivalent (Supplementary Fig. S11). As shown in Fig. S11, while the surface structure itself is the same, the atomic arrangement in the second layer is different between the two surfaces. An earlier ARPES study had suggested that this might affect the electronic structure[20]. The two inequivalent surfaces were revealed in our data when different cleaves showed slightly different FT-STS patterns (Fig. 5, Supplementary Fig. S12 and Fig. S13). Our FT-STS calculations for the second surface, also show that the terminating electronic states are measurably different for the two surfaces. The differences arise from the different weights of bulk and surface states on the two terminations which makes the scattering channels in the measured FT-STS images taken on these two surfaces have different relative strengths. While the match between the calculations and data is not as good for this second surface, both STM as well as theory indicate that the broken inversion symmetry makes the two surfaces electronically inequivalent (Fig. 5).

In conclusion, we have performed a detailed low-temperature STM/STS study on the $MoTe_2$ sample. In the spatially resolved d$I$/d$V$ maps, the quasiparticle interference patterns can be clearly resolved. We thoroughly characterized the electronic structure of $MoTe_2$ with FT-STSs taken at energies both below and above the Fermi level. In the high-resolution FT-STSs above Fermi level, the scattering vectors between the trivial arcs and the projected bulk states can be clearly resolved. Due to the small momentum-space separation between the nontrivial arc and the trivial arc, it is difficult to clearly identify the nontrivial arc features in the FT-STSs. Our data show that the bulk broken inversion symmetry has a measurable effect on the surface electronic properties.

**Methods**

**Sample synthesis.** The $MoTe_2$ single crystals were grown by using a chemical vapor transport method. A mixture of stoichiometric Mo and Te powder were sealed into an evacuated quartz tube with iodine used as a transport agent. The quartz tube was then placed in a double zone furnace with temperature gradient from 900 °C and 800 °C. Large single crystals of centimeter size were obtained after two weeks. The composition and structure of $MoTe_2$ single crystals were checked using x-ray diffraction and an energy-dispersive x-ray spectrometer.

**Scanning tunneling spectroscopy measurements.** High quality $MoTe_2$ samples were cleaved at 77 K and immediately inserted into the STM scanner at 6 K. Differential conductance (d$I$/d$V$) spectra were acquired at 6 K using a standard lock-in technique with ~4 mV$_{rms}$ modulation at a frequency of 987.5 Hz.

**Acknowledgements**

STM work was supported by US Department of Energy, Scanned Probe Division under Award Number DE-SC0014335. S.Y. acknowledges the financial support from Science and Technology Commission of Shanghai Municipality (STCSM) (Grant No. 18QA1403100) and the start-up funding from ShanghaiTech University. The work at Tulane University was supported by the US Department of Energy under Grant No. DE-SC0014208 (support for single crystal growth).

**Author contributions**

S.Y. and V.M. conceived the experiment. Samples were grown by J.H. and obtained from Z.M.. S.Y. and D.I. carried out the STM studies. G.C., T.-R. C and H.L. did all the calculations. S.Y., V.M., and D.I. wrote the paper.

Figure 1:

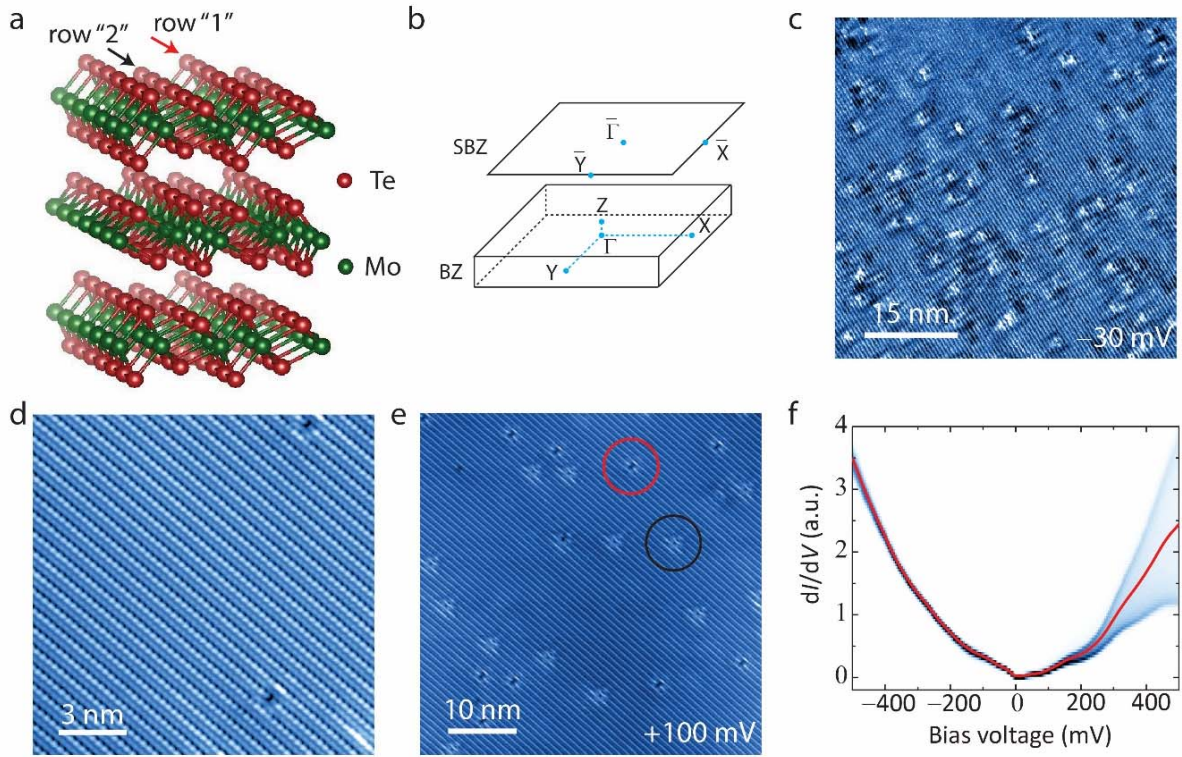

**Figure 1. STM topographies and spatially resolved d$I$/d$V$ spectra on MoTe$_2$.** (**a**) Crystal structure of Td-phase MoTe$_2$. (**b**) Schematics of the surface Brillouin zones (SBZ) and bulk Brillouin zones (BZ) of MoTe$_2$. (**c**) d$I$/d$V$ map at −30 mV. (**d**) High-resolution STM topography taken with +100 mV and 150 pA. (**e**) Large-scale STM topographies taken with +100 mV and 300 pA. The black and red circles indicate two different kinds of atomic impurities on the top Te-layer. (**f**) Spatially-resolved d$I$/d$V$ spectra (blue) taken over a 50 nm by 50 nm area with −100 mV and 750 pA setpoint. The red curve is the averaged d$I$/d$V$ spectrum over this area.

Figure 2:

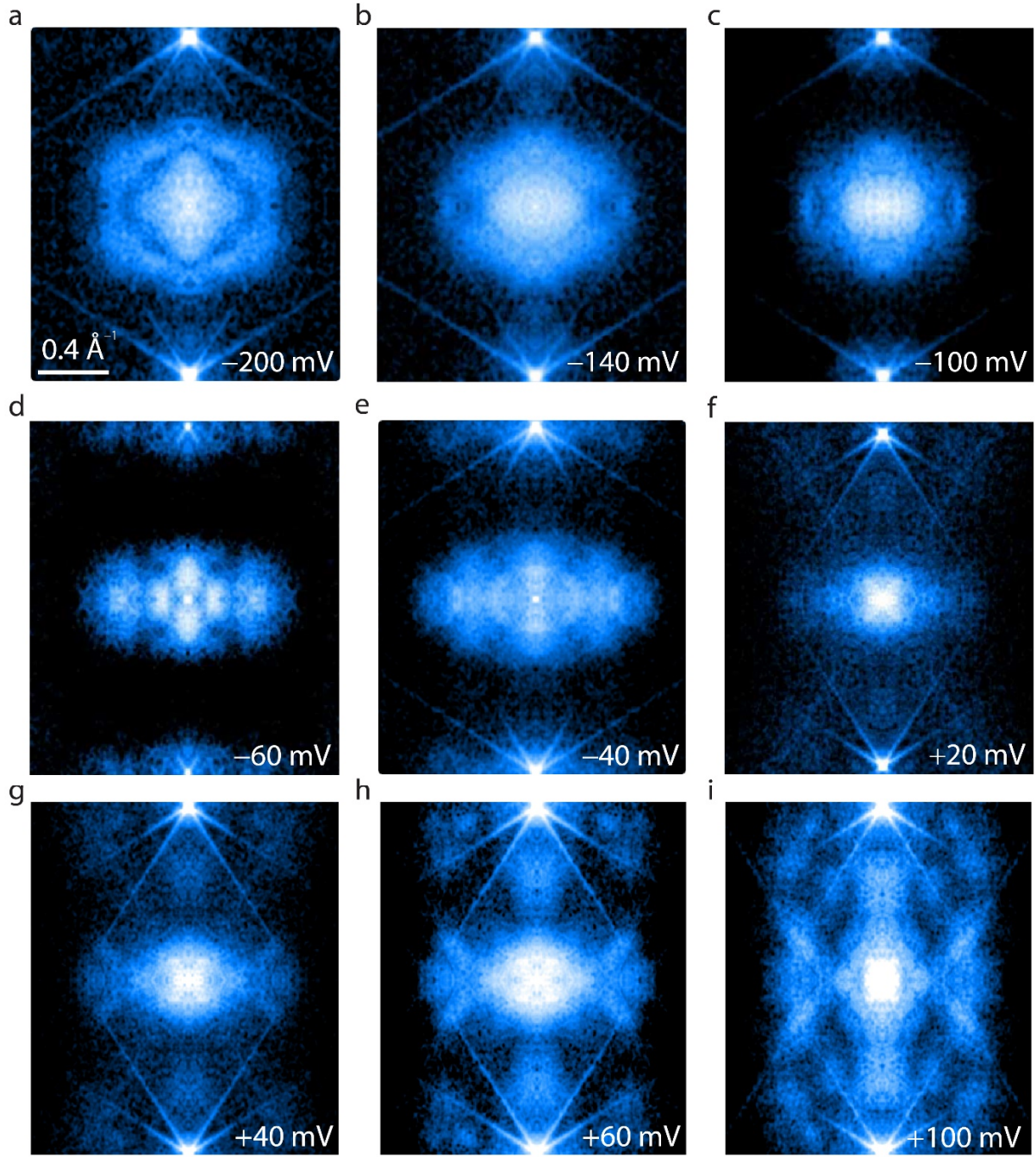

**Figure 2. Quasiparticle interference on MoTe$_2$.** (**a-i**) Mirror-symmetrized FT-STSs at energies of −200 mV, −140 mV, −100 mV, −60 mV, −40 mV, +20 mV, +40 mV, +60 mV and +100 mV respectively. For the FT-STSs with positive sample bias voltages, before performing the Fourier transform, the d*I*/d*V* maps are first normalized with the tunneling current map at +200 mV.

Figure 3:

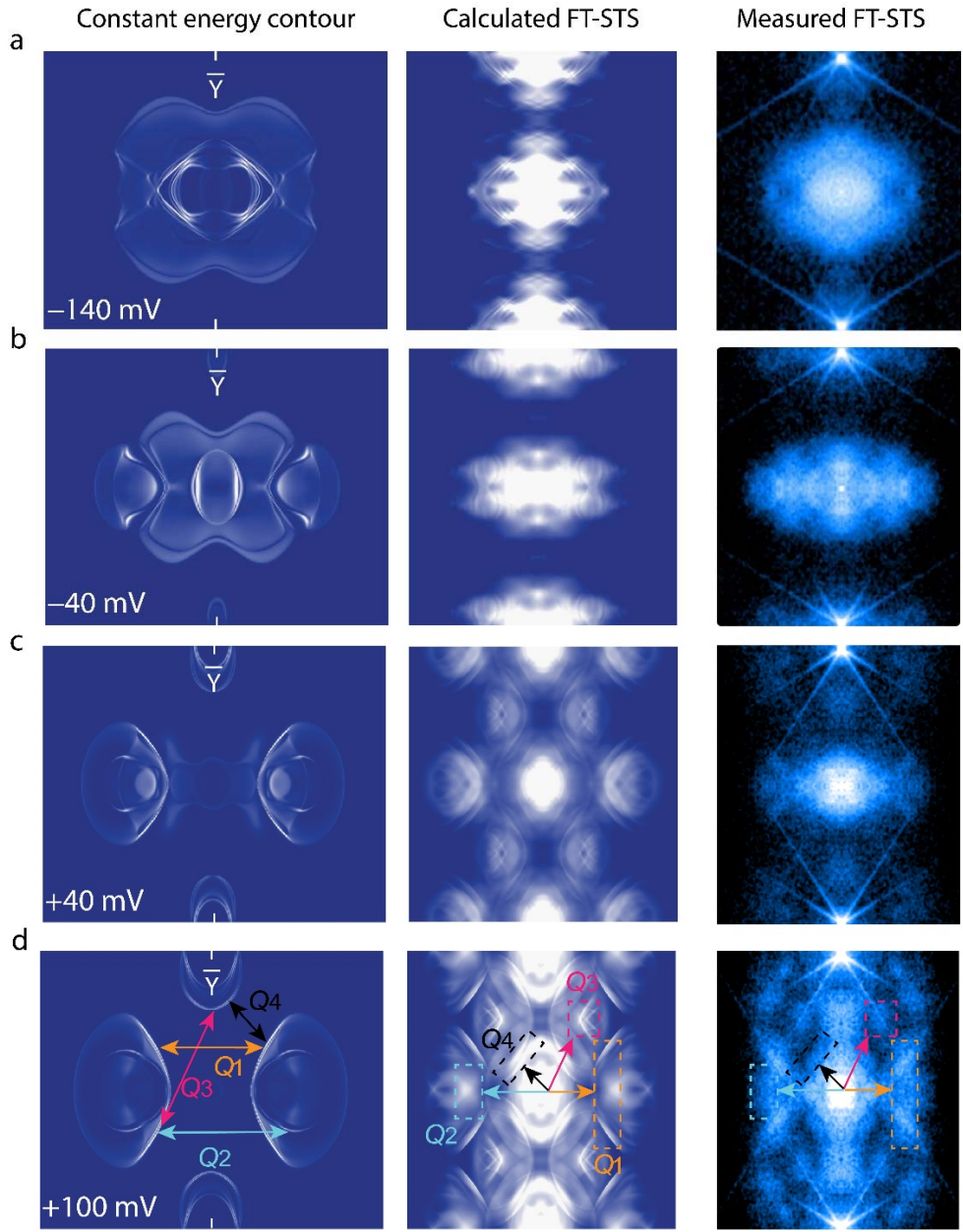

**Figure 3. Comparison between the calculated FT-STSs and the measured FT-STSs.** (**a**) Constant energy contours of MoTe$_2$ at the energy $E = -140$ mV (left panel); theoretically simulated FT-STS at $-140$ mV (middle panel); experimentally measured FT-STS at $-140$ mV (right panel). (**b-c**) The same as **a**, but for energy $E = -40$ mV and $+40$ mV, respectively. The arrows in the right panel of **b** indicate the two short arc-like features. (**d**) The same as **a** to **c**, but for energy $E = +100$ mV. $Q_1$, $Q_2$, $Q_3$ and $Q_4$ are the scattering vectors (see the main text for details).

Figure4:

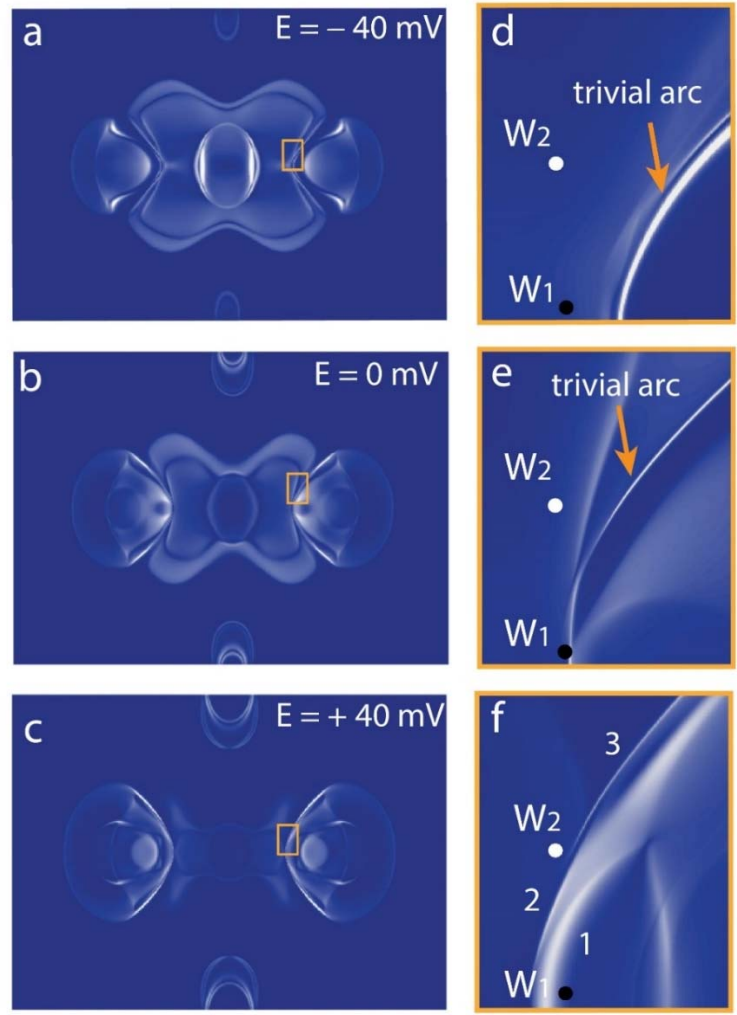

**Figure 4. Trivial Fermi arcs and Weyl nodes in MoTe$_2$.** (**a-c**) Calculated constant energy contours of MoTe$_2$ at the energy $E$ = –40 mV, 0 mV and +40 mV, respectively. (**d-f**) The zoom-in of the orange box in panel **a**, **b** and **c**. The long arc away from the Weyl nodes (black and white dots labelled as W$_1$ and W$_2$) is topologically trivial. The trivial arc indicated by the orange arrows in **d** and **e** is cut into three pieces, labeled as "1'', "2" and "3" in **f**, where "2" is the topological Fermi arc. The size of **d-f** is about 0.08 Å$^{-1}$ by 0.1 Å$^{-1}$.

Figure5:

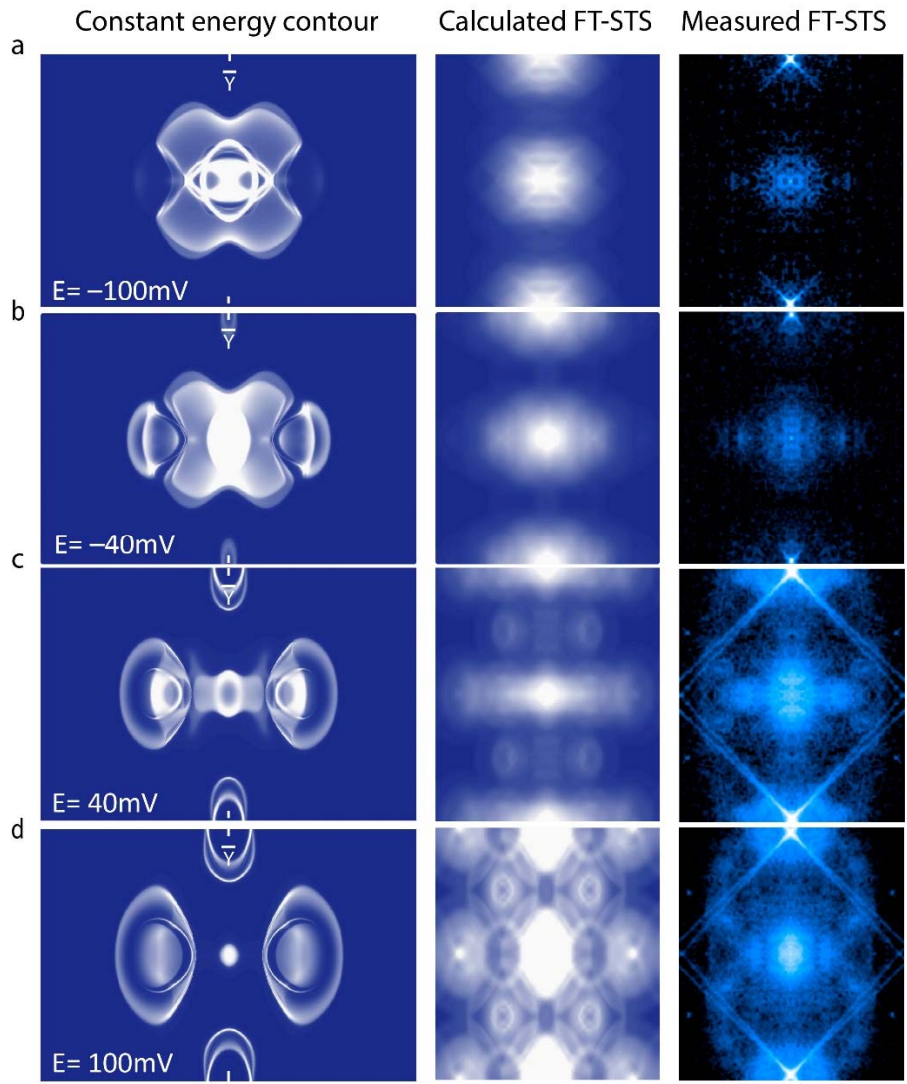

**Figure 5. Calculated FT-STSs and the measured FT-STSs taken on the second surface.** (**a**) Constant energy contours of MoTe$_2$ at the energy $E = -100$ mV (left panel); theoretically simulated FT-STS at $-100$ mV (middle panel); experimentally measured FT-STS at $-100$ mV (right panel). (**b-c**) The same as **a,** but for energy $E = -40$ mV, +40 mV and +100 mV, respectively.